\documentclass{svmult}
 
\usepackage[dvips]{graphicx}
\usepackage{psfrag, amsmath}
\usepackage{makeidx}  
\usepackage{multicol}        
\usepackage[bottom]{footmisc}

\makeindex            

\begin{document}

\title*{Inferring the Composition of a Trader Population in a Financial Market}
\author{Nachi Gupta\inst{1}\and Raphael Hauser\inst{1}\and Neil F. Johnson\inst{2}}
\institute{Oxford University Computing Laboratory, Numerical Analysis Group,\\ Wolfson Building, Parks Road, Oxford OX1 3QD, U.K. \\ \texttt{nachi@comlab.ox.ac.uk}\and Oxford University, Department of Physics, \\Clarendon Building, Parks Road, Oxford OX1 3PU, U.K.}

\maketitle

\section{Introduction}

There has been an explosion in the number of models proposed for understanding and interpreting the dynamics of financial markets.  Broadly speaking, all such models can be classified into two categories:  (a) models which characterize the macroscopic dynamics of financial prices using time-series methods, and (b) models which mimic the microscopic behavior of the trader population in order to capture the general macroscopic behavior of prices.  Recently, many econophysicists  have trended towards the latter by using multi-agent models of trader populations. One particularly popular example is the so-called Minority Game \cite{CMZ2005}, a conceptually simple multi-player game which can show non-trivial behavior reminiscent of real markets. 
Subsequent work has shown that -- at least in principle -- it is possible to train such multi-agent games on real market data in order to make useful predictions \cite{JJ2002,JJH2003,JLJ+2001,AS2005}.
However, anyone attempting to model a financial market using such multi-agent trader games, with the objective of then using the model to make predictions of real financial time-series, faces two problems: (a) How to choose an appropriate multi-agent model? (b) How to infer the level of heterogeneity within the associated multi-agent population?  

This paper addresses the question of how to infer the multi-trader heterogeneity in a market (i.e. question (b)) assuming that the Minority Game, or one of its many generalizations 
\cite{JJ2002,JJH2003}, forms the underlying multi-trader model. We consider the specific case where each agent possesses a pair of strategies and chooses the best-performing one at each timestep. Our focus  is on the uncertainty for our parameter estimates.  Using real financial data for quantifying this uncertainty, represents a crucial step in developing associated risk measures, and for being able to identify pockets of predictability in the price-series.

As such, this paper represents an extension of our preliminary study in  \cite{GHJ2005}. In particular, the present analysis represents an important advance in that it generalizes the use of probabilities for describing the agents' heterogeneity.   Rather than using a probability, we now use a finite measure, which is not necessarily normalized to unit total weight.  This generalization yields a number of benefits such as a stronger preservation of positive definiteness in the covariance matrix for the estimates.  In addition, the use of such a measure removes the necessity to scale the time-series, thereby reducing possible further errors. We also look into the problem of estimating the finite measure over a space of agents which is so large that the estimation technique becomes computationally infeasible.  We  propose a mechanism for making this problem more tractable, by employing many runs with small subsets chosen from the full space of agents.  The final tool we present here is a method for removing bias from the estimates.  As a result of choosing subsets of the full agent space, an individual run can exhibit a bias in its predictions.  In order to estimate and remove this bias, we propose a technique that has been widely used with Kalman Filtering in other application domains.

\section{The Multi-Agent Market Model}

Many multi-agent models -- such as the Minority Game \cite{CMZ2005} and its generalizations 
\cite{JJ2002,JJH2003} -- are based on binary decisions. Agents compete with each other for a limited resource (e.g. a good price) by taking a binary action at each time-step, in response to global price information which is publicly available.  At the end of each time-step, one of the actions is denoted as the winning action.  This winning action then becomes part of the information set for the future.  As an illustration of the tracking scheme, we will use the Minority Game -- however we encourage the reader to choose their own preferred multi-agent game. The game need not be a binary-decision game, but for the purposes of demonstration we will assume that it is. 

\subsection{Parameterizing the Game} 
\label{parestprob}

We provide one possible way of parameterizing the game in order to fit the proposed methodology.  We select a time horizon window of length $T$ over which we score strategies for each agent.  This is a sliding window given by $(w_{k-T}, \ldots, w_{k-1})$ at time step $k$. Here $w_{k} = \mbox{-sgn}(z_{k})$ represents what would have been the winning decision at time $k$, and $z_k$ is the difference in the corresponding price-series, or exchange-rate, $r_k$.
\begin{equation} \label{zkdef}
z_k = r_k - r_{k-1}
\end{equation}
Each agent has a set of strategies which it scores on this sliding time-horizon window at each time-step.  The agent chooses its highest scoring strategy as its winning strategy, and then plays it.  Assume we have $N$ such agents. At each time-step they each play their winning strategy, resulting in a global outcome for the game at that time-step.  Their aggregate actions result in an outcome which we expect to be indicative of the next price-movement in the real price-series. If one knew the population of strategies in use, then one could predict the future price-series with certainty -- apart from a few occasions where ties in strategies might be broken randomly.

The next step is to estimate the heterogeneity of the agent population itself. We choose to use a recursive optimization scheme similar to a Kalman Filter -- however we will also force inequality constraints on the estimates so that we cannot have a negative number of agents of a given type playing the game. Suppose $x_k$ is the vector at time-step $k$ representing the heterogeneity among the $N$ types of agents in terms of their strategies. We can write this as
\begin{equation}
x_k = \left[\begin{array}{c}
x_{1,k} \\
\vdots \\
x_{N,k}
\end{array}\right]
\end{equation}
where we force each element of the vector to be $\geq 0$.
\begin{equation} \label{constraints}
x_{i,k} \geq 0, \forall i
\end{equation}
We provided a very similar scheme recently in \cite{GHJ2005}, in which we had further constrained our estimate to a probability space and as a result also had to re-scale the time-series to better span the interval $[ -1, 1]$.  We now relax the constraint condition (and the re-scaling) since the benefits of lying within the probability space are outweighed by the benefits of allowing the estimate to move out of this space.  One significant benefit of staying within the probability space is the ability to bound covariance matrices on errors (since upper bounds on the probability of certain events are known).  On the other hand, staying constrained to a probability space removes one degree of freedom from our system (i.e. $x_{N,k} = 1-x_{1,k} -\cdots - x_{n-1,k}$).  This can cause the covariance of our estimate to become ill-formed by a possible numerical loss of positive definiteness (which we could prevent by artificially inflating the diagonal elements of the covariance matrix).

\section{Recursive Optimization Scheme} \label {kf}

In the following subsections we introduce the Kalman Filter, after which we will discuss some desirable extensions for our work.

\subsection{Kalman Filter}
A Kalman Filter is a recursive least-squares implementation, which makes only one pass through the data such that it can wait for each measurement to come in real time, and then make an estimate for that time given all the information from the past.  The Kalman Filter holds a minimal amount of information in its memory at each time, yielding a relatively cheap computational cost for solving the optimization problem.  In addition, the Kalman Filter can make a forecast $n$ steps ahead and provides a covariance structure concerning this forecast.  The Kalman Filter is a predictor-corrector system, that is to say,  it makes a prediction and upon observation of real data, it perturbs the prediction slightly as a correction, and so forth.

The Kalman Filter attempts to find the best estimate at every iteration, for a system governed by the following model:
\begin{equation} \label{kfsm} x_{k} = F_{k,k-1} x_{k-1} + u_{k,k-1}, \qquad u_{k,k-1} \sim N(0,Q_{k,k-1}) \end{equation}
\begin{equation} \label{kfmm} z_{k} = H_{k} x_{k} + v_{k}, \qquad v_{k} \sim N(0,R_{k}) \end{equation}
Here $x_{k}$ represents the true state of the underlying system, which in our case is the finite measure over the agents.  $F_{k,k-1}$ is a matrix used to make the transition from state $x_{k-1}$ to $x_{k}$.  In our applications, we choose $F_{k,k-1}$ to be the identity matrix for all $k$ since we assume that locally the finite measure over the strategies doesn't change drastically.  It would be a tough modeling problem to choose another matrix (i.e., not the identity matrix) --   however, if desired we could incorporate a more complex transition matrix into the model, even one that is dependent on previous outcomes.  The variable $z_{k}$ represents the measurement (also called  observation).  $H_{k}$ is a matrix that relates the state space and measurement space by transforming a vector in the state space to the appropriate vector in the measurement space.  For our artificial market model, $H_k$ will be a row vector containing the decisions based on each of the agent's winning strategies.  So $x_k$, the measure over the agents, acts as a weighting on the decisions for each agent, and the inner product $H_k x_k$ can be thought of as a weighted average of the agents' decisions -- this represents the aggregate decision made by the system of agents.  The variables $u_{k,k-1}$ and $v_{k}$ are both noise terms which are normally distributed with mean 0 and variances $Q_{k,k-1}$ and $R_{k}$, respectively.  

The Kalman Filter will at every iteration make a prediction for $x_k$, which we denote by $\hat{x}_{k|k-1}$.  We use the notation ${k|k-1}$ since we will only use measurements provided until time-step $k-1$ in order to make the prediction at time $k$.  We can define the state prediction error $\tilde{x}_{k|k-1}$ as the difference between the true state and the state prediction.
\begin{equation} \label{se1}
\tilde{x}_{k|k-1} = x_{k} - \hat{x}_{k|k-1}
\end{equation}
In addition, the Kalman Filter will provide a state estimate for $x_{k}$, given all the measurements provided up to and including time step $k$.  We denote these estimates by $\hat{x}_{k|k}$.  We can similarly define the state estimate error by
\begin{equation} \label{se2}
\tilde{x}_{k|k} = x_{k} - \hat{x}_{k|k}
\end{equation}
Since we assume $u_{k,k-1}$ is normally distributed with mean $0$, we make the state prediction simply by using $F_{k,k-1}$ to make the transition.  This is given by
\begin{equation} \label{kfsp} \hat{x}_{k|k-1} = F_{k,k-1} \hat{x}_{k-1|k-1} \end{equation}
We can also calculate the associated covariance for the state prediction, which we call the covariance prediction.  This is actually just 
the expectation of the outer product of the state prediction error with itself. This is given by
\begin{equation} \label{kfcp} P_{k|k-1} = F_{k,k-1} P_{k-1|k-1} F_{k,k-1}' + Q_{k,k-1} \end{equation}
Notice that we use the prime notation on a matrix throughout this paper to denote the transpose.  Now we can make a prediction on what we expect to see for our measurement, which we call the measurement prediction, by
\begin{equation} \label{kfmp} \hat{z}_{k|k-1} = H_{k} \hat{x}_{k|k-1} \end{equation}
The difference between our true measurement and our measurement prediction is called the measurement residual, which we calculate by
\begin{equation} \label{kfi} \nu_{k} = z_{k} - \hat{z}_{k|k-1} \end{equation}
We can also calculate the associated covariance for the measurement residual, which we call the measurement residual covariance, by
\begin{equation} \label{kfic} S_{k} = H_{k} P_{k|k-1} H_{k}' + R_{k} \end{equation}

We now calculate the Kalman Gain, which lies at the heart of the Kalman Filter.  This essentially tells us how much we prefer our new observed measurement over our state prediction.  We calculate this by
\begin{equation} \label{kfkg} K_{k} = P_{k|k-1} H_{k}' S_{k}^{-1} \end{equation}
Using the Kalman Gain and measurement residual, we update the state estimate.  If we look carefully at the following equation, we are essentially taking a weighted sum of our state prediction with the Kalman Gain multiplied by the measurement residual.  So the Kalman Gain is telling us how much to `weight in' information contained in the new measurement.  We calculate the updated state estimate by
\begin{equation} \label{kfsu} \hat{x}_{k|k} = \hat{x}_{k|k-1} + K_{k}  \nu_{k} \end{equation}

Finally, we calculate the updated covariance estimate.  This is just the expectation of the outer product of the 
state error estimate with itself.  Here we will 
give the most numerically stable form of this equation, as this form prevents loss of symmetry and best preserves positive definiteness
\begin{equation} \label{kfcu} P_{k|k} = (I - K_{k} H_{k}) P_{k|k-1} (I - K_{k} H_{k})' + K_{k} R_{k} K_{k}^{T} \end{equation}
The covariance matrices throughout the Kalman Filter give us a way to measure the uncertainty of our state prediction, state estimate, 
and the measurement residual.  Also notice that the 
Kalman Filter is recursive, and we require an initial estimate $\hat{x}_{0|0}$ and associated covariance matrix 
$P_{0|0}$.    Here we simply provide the equations of the Kalman Filter without derivation.  For a detailed description of the Kalman Filter, see Ref. \cite{BLK2001}.

\subsection{Nonlinear Equality Constraints} \label{eq_const}

As we are estimating a vector in which each element has a non-negative value, we would like to force the Kalman Filter to have some inequality constraints.  We now introduce a generalization for nonlinear equality constraints followed by an extension to inequality constraints. In particular, 
let's add to our model (Eqs. \eqref{kfsm} and \eqref{kfmm}) the following smooth nonlinear equality constraints
\begin{equation} \label{Lxeq0} e_{k}(x_{k}) = 0 \end{equation}
The constraints provided in Eq. \eqref{constraints} are actually linear.  We present the nonlinear case for further completeness here.  We now rephrase the problem we would like to solve, using the superscript $c$ to denote constrained.  We are given the last prediction and its covariance, the current measurement and its covariance, and a set of equality constraints and would like to make the current prediction and find its covariance matrix.
Let's write the problem we are solving as
\begin{equation} \label{eq1}  z_{k}^{c} = h_{k}^{c}(x_{k}) + v_{k}^{c}, \qquad v_{k}^{c} \sim N(0,R_{k}^{c}) \end{equation}
Here $z_{k}^{c}$, $h_{k}^{c}$, and $v_{k}^{c}$ are all vectors, each having three distinct parts.  The first part will represent the prediction for the current time step, the second part is the measurement, and the third part is the equality constraint.  $z_{k}^{c}$ effectively still represents the measurement, with the prediction treated as a ``pseudo-measurement" with its associated covariance.
\begin{equation} \label{eq2}
z_{k}^{c} = \left[ \begin{array}{c}
	F_{k,k-1} \hat{x}_{k-1|k-1} \\ 
	z_{k} \\
	0
\end{array}\right]
\end{equation}
The matrix $h_{k}^{c}$ takes our state into the measurement space as before
\begin{equation} \label{eq3}
h_{k}^{c}(x_{k}) = \left[ \begin{array}{c} 
	x_{k} \\
	H_{k} x_{k} \\
	e_{k}(x_{k}) 
\end{array} \right]
\end{equation}
Notice that by combining Eqs. \eqref{se1} and \eqref{se2}, we can rewrite the state error prediction as 
\begin{equation} \label {se3}
\tilde{x}_{k|k-1} = F_{k,k-1} \tilde{x}_{k-1|k-1} + u_{k,k-1}
\end{equation}

We can define $v_{k}^{c}$ again as the noise term using Eq. \eqref{se3}.  
\begin{equation} \label{eq4}
v_{k}^{c} = \left[ \begin{array}{c}
	-F_{k,k-1} \tilde{x}_{k-1|k-1} - u_{k,k-1} \\
	v_{k} \\
	0
\end{array} \right]
\end{equation}
$v_{k}^{c}$ will be normally distributed with mean 0 and variance $R_{k}^{c}$.  The diagonal elements of $R_{k}^{c}$ represent the variance of each element of $v_{k}^{c}$.  We define the covariance of the state estimate error at time-step $k$ as $P_{k|k}$.  Notice also that $R_{k}^{c}$ contains no off-diagonal elements.
\begin{equation} \label{eq5}
R_{k}^{c} = \left[ \begin{array}{ccc}
	F_{k,k-1} P_{k-1|k-1} {F_{k,k-1}}' + Q_{k,k-1} & 0 & 0 \\
	0 & R_{k} & 0 \\
	0 & 0 & 0
\end{array} \right]
\end{equation} 
This method of expressing our problem can be thought of as a fusion of (a) the state prediction, and (b) the new measurement at each iteration, under the given equality constraints.  As we did for  the Kalman Filter, we will state the equations here.  The interested reader is referred to Refs. \cite{CWC+2002,WCC2002}.
\begin{equation} \label{xeq}
\hat{x}_{k|k,j} = \left[ \begin{array}{cc}
	0 & I
\end{array} \right] \left[ \begin{array}{cc}
	R_{k}^{c} & H_{k,j}^{c} \\
	{H_{k,j}^{c}}' & 0
\end{array} \right]^{+} \left[ \begin{array}{c}
	z_{k}^{c} - h_{k}^{c}(\hat{x}_{k|k,j-1}^{c}) + H_{k,j}^{c} \hat{x}_{k|k,j-1}^{c} \\
	0
\end{array} \right]
\end{equation}
Throughout this paper, we use the $^{+}$ notation on a matrix to denote the pseudo-inverse.  In this method we are iterating over a dummy variable $j$ within each time-step, 
until we fall within a predetermined convergence bound $\left| \hat{x}_{k|k,j} - \hat{x}_{k|k,j-1} \right| \leq c_k$ or hit a chosen number of maximum iterations.  We initialize our first iteration as $ \hat{x}_{k|k,0} = \hat{x}_{k-1|k-1} $ and use the final iteration as $ \hat{x}_{k|k} = \hat{x}_{k|k,J}$ where $J$ represents the final iteration.
Notice that we allowed the equality constraints to be nonlinear.  As a result, we define $H_{k,j}^{c} = \frac{\partial h_{k}^{c}}{\partial x_{k}}(\hat{x}_{k|k,j-1})$ which gives us a local approximation to the direction of $h_{k}^{c}$.

A stronger form of these equations can be found in Refs. \cite{CWC+2002,WCC2002}, where $R_{k}^{c}$ will reflect the tightening of the covariance for the state prediction based on the new estimate at each iteration of $j$.  We do not use this form and tighten the covariance matrix within these iterations since in the next section we will require the flexibility of changing the number of equality constraints between iterations of $j$.  By not tightening the covariance matrix in this way, we are left with  a larger covariance matrix for the estimate (which shouldn't harm us significantly).  This covariance matrix is calculated as
\begin{equation} \label{Peq}
P_{k|k,j} = - \left[ \begin{array}{cc}
	0 & I
\end{array} \right] \left[ \begin{array}{cc}
	R_{k}^{c} & H_{k,j}^{c} \\
	{H_{k,j}^{c}}' & 0
\end{array} \right]^{+} \left[ \begin{array}{c}
	0 \\
	I
\end{array} \right]
\end{equation}
Notice that for faster computation times, we need only calculate $P_{k|k,j}$ for the final iteration of $j$.  Further, if our equality constraints are in fact independent of $j$, we only need to calculate $H_{k,j}^{c}$ once for each $k$.  This also implies that the pseudo-inverse in Eq. \eqref{xeq} can be calculated only once for each $k$.

This method, while very different from the Kalman Filter presented earlier, provides us with an estimate $\hat{x}_{k|k}$ and a covariance matrix for the estimate $P_{k|k}$ at each time-step, in a similar way to the Kalman Filter.  However, this method allows us to incorporate equality constraints.

\subsection{Nonlinear Inequality Constraints} \label{ineq_const}

We will now extend the equality constrained problem to an inequality constrained problem.  To our system given by equations \eqref{kfsm}, \eqref{kfmm}, and \eqref{Lxeq0}, we will add the smooth inequality constraints given by
\begin{equation} l_{k}(x_{k}) \geq 0. \end{equation}
Our method will be to keep a subset of the inequality constraints active at any time.  An active constraint is simply a constraint that we treat as an equality constraint.  We will ignore any inactive constraint when solving our optimization problem.  After solving the problem, we then check if our solution lies in the space given by the inequality constraints.  If it doesn't, we start from the solution in our previous iteration and move in the direction of the new solution until we hit a set of constraints.  For the next iteration, this set of constraints will be the new active constraints.

We formulate the problem in the same way as before, keeping Eqs. \eqref{eq1}, \eqref{eq2}, \eqref{eq4}, and \eqref{eq5} the same to set up the problem.  However, we replace Eq. \eqref{eq3} by
\begin{equation}
h_{k}^{c}(x_{k}) = \left[ \begin{array}{c} 
	x_{k} \\
	H_{k} x_{k} \\
	e_{k}(x_{k}) \\
	l_{k,j}^{a}(x_{k})
\end{array} \right]
\end{equation}
$l_{k,j}^{a}$ represents the set of active inequality constraints.  Although we keep Eqs. \eqref{eq2}, \eqref{eq4}, and \eqref{eq5} the same, these will need to be padded by additional zeros appropriately to match the size of $l_{k,j}^{a}$.  Now we solve the equality constrained problem consisting of the equality constraints and the active inequality constraints (which we treat as equality constraints) using Eqs. \eqref{xeq} and \eqref{Peq}.  Let's now call the solution from Eq. \eqref{xeq} $\hat{x}_{k|k,j}^{*}$ since we have not yet checked if this solution lies in the inequality constrained space.  In order to check this, we find the vector that we moved along to reach $\hat{x}_{k|k,j}^{*}$.  This is simply 
\begin{equation} d = \hat{x}_{k|k,j}^{*} - \hat{x}_{k|k,j-1} \end{equation}
We now iterate through each of our inequality constraints, to check if they are satisfied.  If they are all satisfied, we choose $t_{\max}=1$. If they are not, we choose the largest value of $t_{\max}$ such that $\hat{x}_{k|k,j-1} + t_{\max} d$ lies in the inequality constrained space.  We choose our estimate to be
\begin{equation} \hat{x}_{k|k,j} = \hat{x}_{k|k,j-1} + t_{\max} d \end{equation}
We also would like to remember the inequality constraints which are being touched in this new solution.  These constraints will now become active for the next iteration and lie in $l_{k,j+1}^{a}$.  Note that $l_{k,0}^{a} = l_{k-1,J}^{a}$, where $J$ represents the final iteration of a given time-step.
We do not perturb the error covariance matrix from Eq. \eqref{Peq} in any way.  Under the assumption that our model is a well-matched model for the data, enforcing inequality constraints (as dictated by the model) should only make our estimate better.  Having a slightly larger covariance matrix is better than having an overly optimistic one based on a bad choice for the perturbation \cite{SS2003}.  In the future, we will investigate how to perturb this covariance matrix correctly.

In our application, our constraints are only to keep each element of our measure positive. Hence we have no equality constraints -- only inequality constraints.  However, we needed to provide the framework to work with equality constraints before we could make the extension to inequality constraints.

Figure \ref{schematic} provides a schematic diagram showing how this optimization scheme fits into the multi-agent game for making predictions.
\begin{figure}[h!] \label{schematic}
\begin{center}
\includegraphics[width=\textwidth]{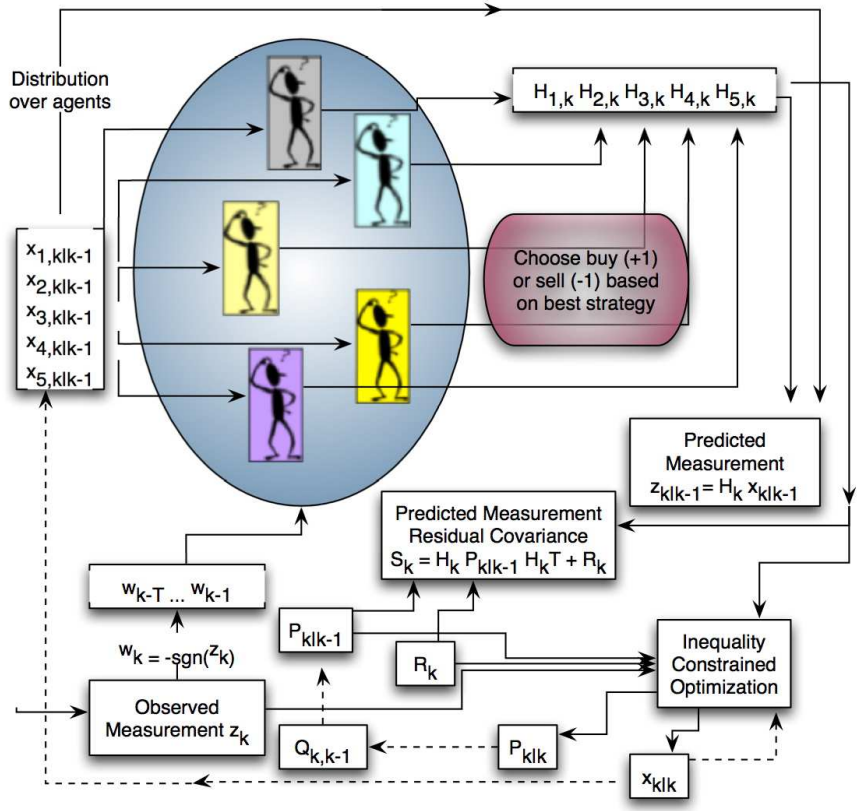}
\end{center}
\caption{Summary of  the recursive method for predicting the heterogeneity of the multi-agent population. We have dropped the $\hat{}$ notation.  Shown is a situation with 5 types of agents, where each type has more than one strategy.  They each score their strategies over the sliding time-horizon window $(w_{k-T} \cdots w_{k-i})$ and choose the best one.  $H_k$ represents the decisions they each make in this time-step, which in the case of a binary-decision game is +1 or -1.  Taking the dot product of the frequencies over the agents and their decisions, we arrive at our prediction for the measurement.  We then allow the recursion into the optimization technique. Since we chose $F_{k,k-1}$ as the identity matrix for all $k$, we omitted it entirely from this diagram.  We also assume initial conditions are provided.  In the next subsection, we describe how we arrive at the noise parameters $Q_{k,k-1}$ and $R_k$, which appear in the diagram.}
\label{schematic}
\end{figure}

\subsection{Noise Estimation} \label{covmatch}

In many applications of Kalman Filtering, the process noise $Q_{k,k-1}$ and measurement noise $R_{k}$ are known.  However, in our application we are not provided with this information a priori so we would like to estimate it.  This can often be difficult to approximate, especially when there is a known model mismatch.  We will present one possible method here which matches the process noise and measurement noise to the past measurement residual process  \cite{Maybeck1982}.
We estimate $R_{k}$ by taking a window of size $W_{k}$ (which is picked in advance for statistical smoothing) and time-averaging the measurement noise covariance based on the measurement residual process and the past states.  If we refer back to Eq. \eqref{kfic}, we can simply calculate this by
\begin{equation} \label{Rk}
\hat{R}_{k} = \frac{1}{W_{k}-1} \sum_{j=k-W_{k}}^{k-1}{\nu_{j} {\nu_{j}}' - H_{j} P_{j|j-1} H_{j}'}
\end{equation}
We can now use our choice of $R_{k}$ along with our measurement residual covariance $S_{k}$, to estimate $Q_{k,k-1}$.  Combining Eqs. \eqref{kfcp} and \eqref{kfic} we have
\begin{equation}
S_{k} = H_{k} (F_{k,k-1} P_{k-1|k-1} {F_{k,k-1}}' + Q_{k,k-1}) {H_{k}}' + R_{k}
\end{equation}
Bringing all $Q_{k,k-1}$ terms to one side leaves us with
\begin{equation}
H_{k} Q_{k,k-1} {H_{k}}' = S_{k} - H_{k} F_{k} P_{k-1|k-1} {F_{k}}' {H_{k}}' - R_{k}
\end{equation}
Solving for $Q_{k,k-1}$ gives us
\begin{equation} \label{Qk}
\hat{Q}_{k,k-1} = {\left({H_{k}}' H_{k}\right)}^{+} {H_{k}}'  \left( S_{k} - H_{k} F_{k} P_{k-1|k-1} {F_{k}}' {H_{k}}' - R_{k} \right) H_{k} {\left({H_{k}}' H_{k}\right)}^{+}
\end{equation}

Note that it may be desirable to keep $\hat{Q}_{k,k-1}$ diagonal if we do not believe the process noise has any cross-correlation.  It is rare that one would expect a cross-correlation in the process noise.  In addition, keeping the process noise diagonal has the effect of making our covariance matrix `more positive definite'.  This can be done simply by setting the off-diagonal terms of $\hat{Q}_{k,k-1}$ equal to $0$. It is also important to keep in mind that we are estimating covariance matrices here which must be symmetric and positive semidefinite, and the diagonal elements should always be greater than or equal to zero since these are variances.

\section{Estimation in the Presence of an Ecology of Many Agent Types}

It is very likely that we will come across multi-agent markets, and hence models, with many different agent types, e.g., $N > 100$.  As $N$ grows, not only does our state space grow linearly, but our covariance space will grow quadratically.  We quickly reach areas where we may no longer be in a computationally feasible region. For example, if we look at strategies for agents playing the Minority Game and define a type of agent to have exactly 2 strategies, we see that as we increase the memory sizes of our agents our full set of pairs of strategies grows very quickly in relation to the memory size (e.g., $m=1$ yields $2^{2^1}=4$ strategies and $\binom{4}{2} = 6$ pairs of strategies, $m=2$ yields $2^{2^2}=16$ strategies and $\binom{16}{2} = 120$ pairs of strategies, $m=3$ yields $2^{2^3}=256$ strategies and $\binom{256}{2} = 32640$ pairs of strategies, $m=4$ yields $2^{2^4}=65536$ strategies and $\binom{65536}{2} = 2147450880$ pairs of strategies, \ldots).  If we were interested in simultaneously allowing all possible pairs of strategies, our vectors and matrices for these computations would have a dimension that would not be of reasonable complexity, especially in situations where real-time computations are needed.
In such situations, we propose selecting a subset of the full set of strategies uniformly at random, and choosing these as the only set that could be in play for the time-series.  We can then do this a number of times and average over the predictions and their covariances.  We would hope that this would cause a smoothing of the predictions and remove outlier points.  In addition we might notice certain periods that are generally more predictable by doing this, which we call {\em pockets of predictability}.

\subsection{Averaging over Multiple Runs} \label{mce}

For each run $j$ of our $M$ runs, we have our predicted measurement at time $k$ given by $\hat{z}_{k,j}$ and our predicted covariance for the measurement residual as $S_{k,j}$.  Using the predicted measurements, we can simply average to find our best estimate of the prediction.
\begin{equation} \label{mc1}
\hat{z}^*_{k} = \sum_{j=1}^M \frac{\hat{z}_{k,j}}{M}
\end{equation}
Similarly, we can calculate our best estimate of the predicted covariance for the measurement residual:
\begin{equation}
S^*_{k} = \sum_{j=1}^M \frac{S_{k,j}}{M}
\end{equation}

It is important to note here that since $\hat{z}^*_{k}$ and $S^*_{k}$ are both estimators, as $M$ tends to $\infty$, we expect the standard error of the mean for both to tend towards 0.  Also, note that we chose equal weights when calculating the averages; we could have alternatively chosen to use non-equal weights had we developed a system for deciding on the weights.  

\subsection{Bias Estimation}

Since we are choosing subsets of the full strategy space, we expect that in some runs a number of the strategies might tend to behave in the same way.  This doesn't mean that the run is useless and provides no information.  In fact, it could be the case that the run provides much information -- it is just that the predictions always tend to be biased in one direction or the other.  So what we might like to do is remove bias from the system.
The simplest way to do this is to augment the Kalman Filter's state space with a vector of elements representing possible bias \cite{Friedland1969}.  We can model this bias as lying in the state space, the measurement space, or some combination of elements of either or both.  We redefine the model for our problem as 
\begin{equation} \label{bkfsm} x^b_{k} = F^b_{k,k-1} x^b_{k-1} + u_{k,k-1}, \qquad u_{k,k-1} \sim N(0,Q^b_{k,k-1}) \end{equation}
\begin{equation} \label{bkfmm} z_{k} = H^b_{k} x^b_{k} + v_{k}, \qquad v_{k} \sim N(0,R_{k}) \end{equation}
where $x^b_k$ represents the augmented state and $b_k$ is the bias vector at time step $k$
\begin{equation}
x^b_k = \left[ \begin{array}{c}
x _k\\
b_k
\end{array} \right]
\end{equation}
The transition matrix must also be augmented to match the augmented state.  In the top left corner, we place our original transition matrix, and in the top right corner we place $B_{k,k-1}$ representing how the estimated bias term should be added into the dynamics.  In the bottom left we have the zero matrix so the bias term is not dependent on the state $x_k$, and in the bottom right we have the identity matrix indicating that the bias is updated by itself exactly at each time.
\begin{equation} \label{biasphi}
F^b_{k,k-1} = \left[\begin{array}{cc}
F_{k,k-1} & B_{k,k-1} \\
0 & I
\end{array} \right]
\end{equation}
Similarly, we horizontally augment our measurement matrix, where $C_k$ represents how the bias terms should be added into the measurement space.
\begin{equation} \label{biash}
H^b_{k} = \left[\begin{array}{cc}
H_{k} & C_k
\end{array} \right]
\end{equation}
For the process noise, we keep the off diagonal elements as 0, assuming no cross-correlations between the state and the bias.  We also generally assume no noise in the bias term and keep its noise covariance as 0, as well. Of course, this can be changed easily enough if the reader would like to model the bias with some noise.
\begin{equation}
Q^b_{k,k-1} = \left[\begin{array}{cc}
Q_{k,k-1} & 0 \\
0 & 0
\end{array} \right]
\end{equation}
We can take this bias model framework and place it into the inequality constrained filtering scheme provided earlier, with the model given by Eq. \eqref{eq1}, where we simply use the augmented states when necessary, rather than the regular state space (e.g. let $x_k = x^b_k$).

\section{Example for a Foreign-Exchange Rate Series}

We now apply these ideas to a data set of hourly USD/YEN foreign exchange rate data, from 1993 to 1994, using the Minority Game.  We do not claim that this is a good model for the time-series, but it does contain some of the characteristics we might expect to see in this time-series.  We look at all pairs of strategies with memory size $m=4$ of which there are 2,147,450,880 as our set of possible types for agents.  Since the size of this computation would not be tractable, we take a random subset of 5 of these types and use these 5 as the only possible types of agents to play the game. We perform 100 such runs, each time choosing 5 types at random.
In addition, we allow for a single bias removal term.  We could have many more terms for the bias, but we only use 1 in order to limit the growth of the state space.  We assume the entire bias lies in a shifting of the measurements, so we don't use a $B_{k,k-1}$ from Eq. \eqref{biasphi} and only choose $C_k$ in Eq. \eqref{biash} to be the identity matrix  -- or in our case simply 1 since $C_k$ is 1x1.

\begin{figure}[h!]
\begin{center}
\psfrag{Hourly USD/YEN FX-Rate from 1993 to 1994}{Hourly USD/YEN FX-Rate from 1993 to 1994}
\psfrag{Residual Log Return}{Residual Log Return}
\psfrag{Time}{Time}
\psfrag{Residuals}{Residuals}
\psfrag{+/- 3 Standard Deviations}{$\pm 3$ Std. Deviations}
\psfrag{Standard Error of the Mean}{Std. Error of the Mean}
\includegraphics[width=\textwidth]{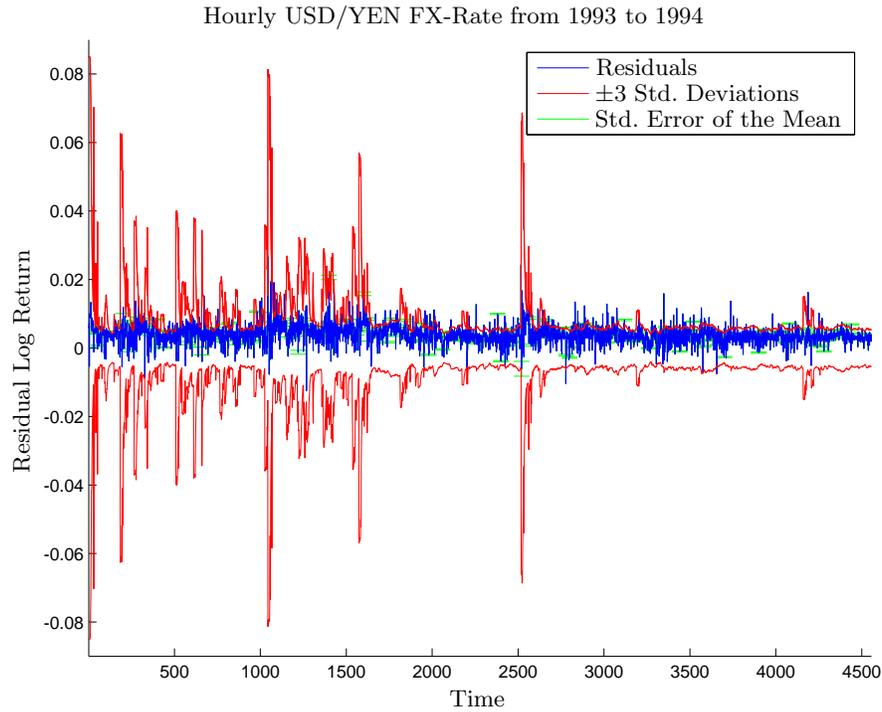}
\end{center}
\caption{We show the residuals of the log returns plotted with 3$\sigma$'s centered about 0 and the standard error of the mean over the 100 runs.  Despite the one parameter bias removal, we still see a general bias in the data without which the residuals look much cleaner.  Perhaps a more complex bias model would remove this.}
\label{fig}
\end{figure}

For the analysis of how well our forecasts perform, we calculate the residual log returns and plot these.  Given our time-series, we can calculate the log return of the exchange rate as $l_k = \log(r_k) - \log(r_{k-1})$.  Note that based on our definition for $z_k$ from Eq. \eqref{zkdef}, we can write the log return also as $l_k = \log(r_{k-1}+z_k) - \log(r_{k-1})$.  Similarly, we can define our predicted forecast for the log return as $\hat{l}_k = \log(r_{k-1}+\hat{z}_k) - \log(r_{k-1})$.  Given these two quantities, we can calculate the residual of the predicted log return and the observed log return as $\tilde{l}_k = l_k - \hat{l}_k$.  Using the delta method \cite{CB2002}, we can also calculate the variance of this residual to be $\frac{S_k}{(r_{k-1})^2}$. We perform 100 such runs, which we average over using the method described in Section \ref{mce}.  A good test for whether our variances are overly optimistic is to check if the measure satisfies the Chebyshev Inequality.  For example, we can check visually that no more than about $\frac{1}{9}$ of the residuals lie within $3\sigma$'s.  We plot the residual log return along with 3$\sigma$ bounds and the standard error of the mean in Figure \ref{fig}.  Visually, we would say the residuals in Figure \ref{fig} would certainly satisfy this condition if they were centered about 0. Maybe further bias removal would readily achieve this. 

\section{Conclusion}

This paper has looked at how one can infer the multi-trader
heterogeneity in a financial market. The market itself could be an
artificial one (i.e., simulated, like the so-called Minority Game), or a real one -- the technique is essentially the same.  The
method we presented provides a significant extension of previous work
 \cite{GHJ2005}. When coupled with an underlying market model
that better suits the time-series under analysis, these techniques
could provide useful insight into the composition of multi-trader
populations across a wide range of markets. We have also provided a
framework for dealing with markets containing a very large number of
active agents. Together, these ideas can yield superior prediction
estimates of a real financial time-series.


\bibliographystyle{splncs}
\bibliography{kolkatarefs}
\printindex
\end{document}